# Building a PubMed knowledge graph

Jian Xu[1], Sunkyu Kim[2], Min Song[3], Minbyul Jeong[2], Donghyeon Kim[2], Jaewoo Kang[2], Justin F. Rousseau[4], Xin Li[5], Weijia Xu[6], Vetle I. Torvik[7], Yi Bu[8], Chongyan Chen[5], Islam Akef Ebeid[5], Daifeng Li[1], & Ying Ding[4,5]

[1]School of Information Management, Sun Yat-sen University, Guangzhou, China.
[2]Department of Computer Science and Engineering, Korea University, Seoul, South Korea.
[3]Department of Library and Information Science, Yonsei University, Seoul, South Korea.
[4]Dell Medical School, University of Texas at Austin, Austin, TX, USA.
[5]School of Information, University of Texas at Austin, Austin, TX, USA.
[6]Texas Advanced Computing Center, Austin, TX, USA.
[7]School of Information Sciences, University of Illinois at Urbana-Champaign, Champaign, IL, USA.
[8]Center for Complex Networks and Systems Research, Luddy School of Informatics, Computing, and Engineering, Indiana University, Bloomington, IN, USA.

Corresponding authors: Ying Ding (ying.ding@austin.utexas.edu), Daifeng Li (lidaifeng@mail.sysu.edu.cn)

## Abstract

PubMed® is an essential resource for the medical domain, but useful concepts are either difficult to extract or are ambiguated, which has significantly hindered knowledge discovery. To address this issue, we constructed a PubMed knowledge graph (PKG) by extracting bio-entities from 29 million PubMed abstracts, disambiguating author names, integrating funding data through the National Institutes of Health (NIH) ExPORTER, collecting affiliation history and educational background of authors from ORCID®, and identifying fine-grained affiliation data from MapAffil. Through the integration of the credible multi-source data, we could create connections among the bio-entities, authors, articles, affiliations, and funding. Data validation revealed that the BioBERT deep learning method of bio-entity extraction significantly outperformed the state-of-the-art models based on the F1 score (by 0.51%), with the author name disambiguation (AND) achieving a F1 score of 98.09%. PKG can trigger broader innovations, not only enabling us to measure scholarly impact, knowledge usage, and knowledge transfer, but also assisting us in profiling authors and organizations based on their connections with bio-entities. The PKG is freely available on Figshare (https://figshare.com/s/6327a55355fc2c99f3a2, simplified version that exclude PubMed raw data) and TACC website (http://er.tacc.utexas.edu/datasets/ped, full version).

## Background and Summary

Experts in healthcare and medicine communicate in their own languages, such as SNOMED CT, ICD-10, PubChem, and gene ontology. These languages equate to gibberish for laypeople, but for medical minds, they are an intricate method of transporting important semantics and consensus capable of translating diagnoses, medical procedures, and medications among millions of physicians, nurses, and medical researchers, thousands of hospitals, hundreds of pharmacies, and a multitude of health insurance companies. These languages (e.g., genes, drugs, proteins, species, and mutations) are the backbone of quality healthcare. However, they are deeply embedded in publications, making literature searches increasingly onerous because conventional text mining tools and algorithms continue to be ineffective. Given that medical domains are deeply divided, locating collaborators across domains is arduous. For instance, if a researcher wants to study ACE2 gene related to COVID-19, he or she would like to know the following: which researchers are currently actively studying ACE2 gene, what are the related genes, diseases, or drugs discussed in these articles related to ACE2 gene, and with whom could the researcher collaborate? This is a strenuous position to be in, and the aforementioned problems diminish the curiosity directed at the topic.

Many studies have been devoted to building open-access datasets to solve bio-entity recognition problems. For example, Kai *et al.*[1] used a conditional random field classifier-based tool to recognize the named entities from PubMed and PubMed Central. Bell *et al.*[2] performed a large-scale integration of a diverse set of bio-entities and their relationships from both bio-entity datasets and PubMed literature. Although these open-access datasets are predominantly about bio-entity recognition, researchers have also been interested in extracting other types of entities and relationships in PubMed, including the mapping of author affiliations to cities and their geocodes[3,4], author name disambiguation[5] (AND), and author background information collections[6]. Although the focus of previous research has been on limited types of entities, the goal of our study was to integrate a comprehensive dataset by capturing bio-entities, disambiguated authors, funding, and fine-grained affiliation information from PubMed literature.

Figure 1 illustrates the bio-entity integration framework. This framework consists of two parts: (1) bio-entity extraction, which contains entity extraction, named entity recognition (NER), and multi-type normalization, and (2) integration, which connects authors, ORCID, and funding information.

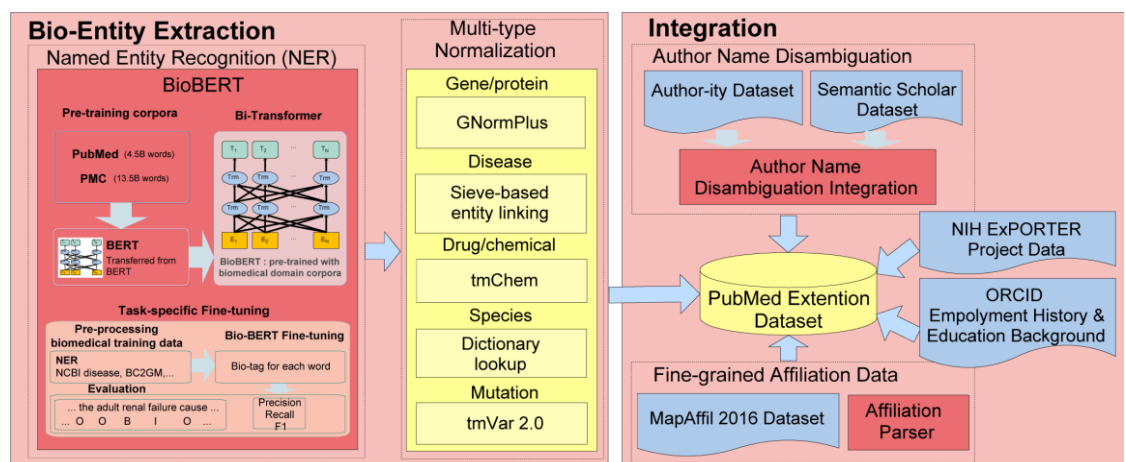

Figure 1. Bio-entity integration framework for PKG

The process illustrated in Figure 1 can be described as follows. First, we applied the high-performance deep learning method Bidirectional Encoder Representations from Transformers for Biomedical Text Mining (BioBERT) [7,8] to extract bio-entities from 29 million PubMed abstracts. Based on the evaluation, this method significantly outperformed the state-of-the-art methods based on the F1 score (by 0.51%, on average). Then, we integrated two existing high-quality author disambiguation datasets: Author-ity[5] and Semantic Scholar[9]. We obtained the disambiguated authors of PubMed articles with full coverage and quality of 98.09% in terms of the F1 score. Next, we integrated additional fields from credible sources into our dataset, which included the projects funded by the National Institutes of Health (NIH) [10], the affiliation history and educational background of authors from ORCID[6], and fine-grained region and location information from the MapAffil 2016 dataset[11]. We named this new interlinked dataset "PubMed Knowledge Graph" (PKG). PKG is by far the most comprehensive, up-to-date, high-quality dataset for PubMed regarding bio-entities, articles, scholars, affiliations, and funding information. Being an open dataset, PKG contains rich information ready to be deployed, facilitating the effortless development of applications such as finding experts, searching bio-entities, analyzing scholarly impacts, and profiling scientist's careers.

## Methods

**Bio-Entity Extraction**

The bio-entity extraction component has two models: (1) an NER model, which recognizes the named entities in PubMed abstracts based on the BioBERT model[7], and (2) a multi-type normalization model, which assigns unique IDs to recognize biomedical entities.

**(1) Named Entity Recognition (NER).** The NER task recognizes a variety of domain-specific proper nouns in a biomedical corpus and is perceived as one of the most notable biomedical text mining tasks. In contrast to previous studies that have built models based on long short-term memory (LSTM) and conditional random fields (CRFs)[12,13], the recently proposed Bidirectional Encoder Representations from Transformers (BERT)[14] model achieves excellent performance for most of the NLP tasks with minimal task-specific architecture modifications. The transformers applied in BERT connect the encoders and decoders through self-attention for greater parallelization and reduced training time. BERT was designed as a general-purpose language representation model that was pre-trained on English Wikipedia and BooksCorpus. Consequently, it is incredibly challenging to maintain high performance when applying BERT to biomedical domain texts that contain a considerable number of domain-specific proper nouns and terms (e.g., BRCA1 gene and Triton X-100 chemical). BERT required refinement, so BioBERT—a neural network-based high-performance NER model—was developed. Its purpose is to recognize the known biomedical entities and discover new biomedical entities.

First, in the NER component, the case-sensitive version of BERT is used to initialize BioBERT. Second, PubMed articles and PubMed Central articles are used to pre-train BioBERT's weights. The pre-trained weights are then fine-tuned for the NER task. While fine-tuning BERT (BioBERT), we use WordPiece tokenization[15] to mitigate the out-of-vocabulary issue. WordPiece embedding is a method of dividing a word into several units (e.g., Immunoglobulin divided into I ##mm##uno ##g ##lo ##bul ##in) and expressing each unit. This technique is effective at extracting the features associated with uncommon words. The NER models available in BioBERT can predict the following seven tags: IOB2 tags (i.e., Inside, Outside, and Begin)[16], X (i.e., a sub-token of WordPiece), [CLS] (i.e., the leading token of a sequence for classification), [SEP] (i.e., a sentence delimiter), and PAD (i.e., a padding of each word in a sentence). The NER models were fine-tuned as follows[8]:

$$p(T_i) = softmax(T_i W^T + b)_k, \qquad k = 0,1,\dots,6 \quad (1)$$

where $k$ represents the indexes of seven tags {B, I, O, X, [CLS], [SEP], PAD}, $p$ is the probability distribution of assigning each $k$ to token $i$, and $T_i \in R^H$ is the final hidden representation, which is calculated by BioBERT for each token $i$. $H$ is the hidden size of $T_i$, $W \in R^{K \times H}$ is a weight matrix between $k$ and $T_i$, $K$ represents the number of tags and is equal to 7, and $b$ is a K-dimensional vector that records the bias on each $k$. The classification loss $L$ is calculated as follows:

$$L(\Theta) = -\frac{1}{N}\sum_{i=1}^{N} log(p(y_i|T_i;\Theta) \quad (2)$$

where $\Theta$ represents the trainable parameters, and $N$ is the sequence length.

First, a tokenizer is applied to words in a sentence on a dataset with labels in the CoNLL format[17]. The WordPiece algorithm is then applied to the sub-words of each word. Consequently, BioBERT is able to extract diverse types of bio-entities. Furthermore, an entity or two entities with frequently-occurring token interaction will be marked with more than one entity type span (26.2% for all PubMed abstracts). Based on the calculated probability distribution, we are able to choose the correct entity type when entities are tagged with more than two types according to the probability-based decision rules[8].

**(2) Multi-Type Normalization.** Because an entity may be referred to by several synonymous terms (synonyms), and a term can be polysemous if it refers to multiple entity types (polysemy), we require a normalization process for the extracted entities. However, it is a daunting challenge to build a single normalization tool for multiple entity types because there exist various normalization models that depend on the type of entity. We addressed this issue by combining multiple NER normalization models into one multi-type normalization model that assigns IDs to extracted entities. Table 1 illustrates the statistics of the proposed normalization model.

Table 1. Multi-type normalization model and dictionaries.

| Entity types | Normalization models | Dictionaries | # of IDs | # of names | Avg. # of names per ID |
|---|---|---|---|---|---|
| Gene/Protein | GNormPlus | Entrez Gene[18] | 139,375 | 248,581 | 1.8 |
| Disease | Sieve-based entity linking[19] | MeSH[20], OMIM[21], SNOMED CT[22], PolySearch2[23] | 32,954 | 172,650 | 5.2 |
| Drug/Chemical | tmChem without Ab3P | MeSH[20], ChEBI[24], DrugBank[25], US FDA-approved drugs | 518,223 | 2,571,570 | 5.0 |
| Species | Dictionary lookup | NCBI Taxonomy | 398,037 | 3,119,005 | 7.8 |
| Mutation | tmVar 2.0 | dbSNP[26], Clin Var[27] | 208,474 | 302,498 | 1.5 |
| Total | | | 1,297,063 | 6,414,304 | 4.9 |

The multi-type normalization model is based on a normalization model per entity type (Table 1). To improve the number of normalized entities, we added the disease names from the PolySearch2 dictionary (76,001 names of 27,658 diseases) to the sieve-based entity linking dictionary (76,237 names of 11,915 diseases). We also added the drug names from DrugBank[25] and the U.S. Food and Drug Administration (FDA) to the tmChem dictionary. Because there are no existing normalization model for species, we normalized species based on dictionary lookup. Using tmVar 2.0, we created a dictionary of mutations with normalized mutation names, in which a mutation with several names was assigned to one normalized name or ID.

**Author Name Disambiguation (AND)**

Despite a rigorous effort to create global author IDs (e.g., ORCID and ResearcherID), most articles in PubMed, particularly those before 2003 (the year in which the field ORCID was added into PubMed), provide limited author information with respect to last name, first initial, and affiliation (only for first authors before 2014). Author information is not effective meta-data to be used directly as a unique identifier because different people may have the same names, and the names and affiliations of an individual can change over time. AND is essential for identifying unique authors.

In recent decades, researchers have made several attempts to solve the AND problem, using three types of methods. The first type of method relies on manual matching of articles with authors by surveying scientists or consulting curricula vitae (CVs) gathered from the Internet [28]. Although this type of method ensures high accuracy, a considerable amount of investment in labor is required to collect and code the data, which is impractical for huge datasets. The second type of method uses publicly-accessible registry platforms, such as ORCID or Google Scholar, to help researchers identify their own publications, which produces a source of highly accurate and low-cost accessible disambiguation of authorship for large numbers of authors. However, registries cover only a small proportion of researchers[29,30], which introduces a form of survivor bias into samples. The third type of method uses an automated approach to estimate the similarity of author instance feature combinations and identify whether they refer to the same person. The features for automated AND include author name, author affiliation, article keywords, journal names[31], coauthor information[32], and citation patterns[33]. Automated methods typically rely on supervised or unsupervised machine learning, in which the machine learns how to weigh the various features associated with author names and where to assign a pair of author names either to the same author or to two different authors[34,35]. This type of method can potentially avoid the shortcomings of the previous two types. Moreover, automated methods have been improved to a high level of accuracy after years of development.

For PubMed, automated methods are the optimal choice because they can overcome the shortcomings of the other two methods while simultaneously providing high-quality AND results for the entire dataset. Several scholars have disambiguated the authors using automated methods. Although the evaluations of these results have exhibited different levels of accuracy and coverage limitations, we believe that integrating them with due diligence can yield a high-quality AND dataset with full coverage of PubMed articles.

According to our investigation, a high-quality PubMed AND dataset with complete coverage can be obtained through the integration of the following two existing AND datasets:

(1) Author-ity: The Author-ity database uses diverse information about authors and publications to determine whether two or more instances of the same name (or of highly similar names) on different papers represent the same person. According to the AND evaluation based on the method discussed in the section *Technical Validation*, the F1 score of Author-ity is 98.16%, which is the highest accuracy result that we have observed. However, this dataset only covers authors before 2009.

(2) Semantic Scholar: It trains a binary classifier to merge a pair of author names and use the pair to create author clusters incrementally. According to the AND evaluation based on the method discussed in the section *Technical Validation*, the F1 score of Semantic Scholar is 96.94%, which is 1.22% lower than that of Author-ity. However, it has the most comprehensive coverage of authors.

Because the Author-ity dataset has a higher F1 score than the Semantic Scholar dataset, we selected the author's unique ID of the Author-ity dataset as the primary AND_ID. AND_ID is limited by time range (containing PubMed papers before 2009); however, we supplemented authors after 2009 using the AND result from Semantic Scholar. The following steps were applied:

Step 1: Allocate the author's unique ID to each author instance according to the Author-ity AND results such that authors from the Author-ity dataset (before 2010) have unique author IDs.

Step 2: For authors that have the same Semantic Scholar AND_ID but never appear in the Author-ity dataset, we generated a new AND_ID to label them. For example, author "Pietranico R." published two papers in 2012 and 2013 and had two corresponding author instances. Because all papers that "Pietranico R." published were after 2009, they were not covered by Author-ity and therefore had no AND_ID allocated by Author-ity. However, the authors disambiguated correctly by Semantic Scholar were allocated unique AND_IDs in Semantic Scholar. To maintain the consistency in labeling, we generated a new AND_ID continuing AND-IDs of Author-ity to label these two author instances as disambiguated by Semantic Scholar.

Step 3: For author instances with a unique AND_ID in Semantic Scholar and in which authors (at least one) had the same Author-ity AND_ID, we allocated the Author-ity AND_ID to all author instances as their unique ID. For example, "Maneksha S." published three papers in 2007, 2009, and 2010, and the first two author instances had unique Author-ity AND_ID. However, the last one had no Author-ity AND_ID because it was beyond the time coverage of the Author-ity dataset. Nevertheless, based on the AND results of Semantic Scholar, the three author instances had an identical AND_ID. Therefore, the last author instance with no Author-ity AND_ID could be labeled with the same ID as the other two author instances.

**Extended Multi-source Information Integration**

In addition to bio-entity extraction by BioBERT and AND, we made a considerable effort to integrate PubMed by extending multi-source data into PKG, which exploited the mapping connections between AND_ID and the PubMed identifier (PMID) to build relationships between different objects to provide a comprehensive overview of the PubMed dataset. These integrated data includes the funding data from NIH ExPORTER, the affiliation history and educational background of authors from ORCID, and the fine-grained region and location information from the MapAffil 2016 dataset. The entities and their associated relationships are depicted in Figure 2.

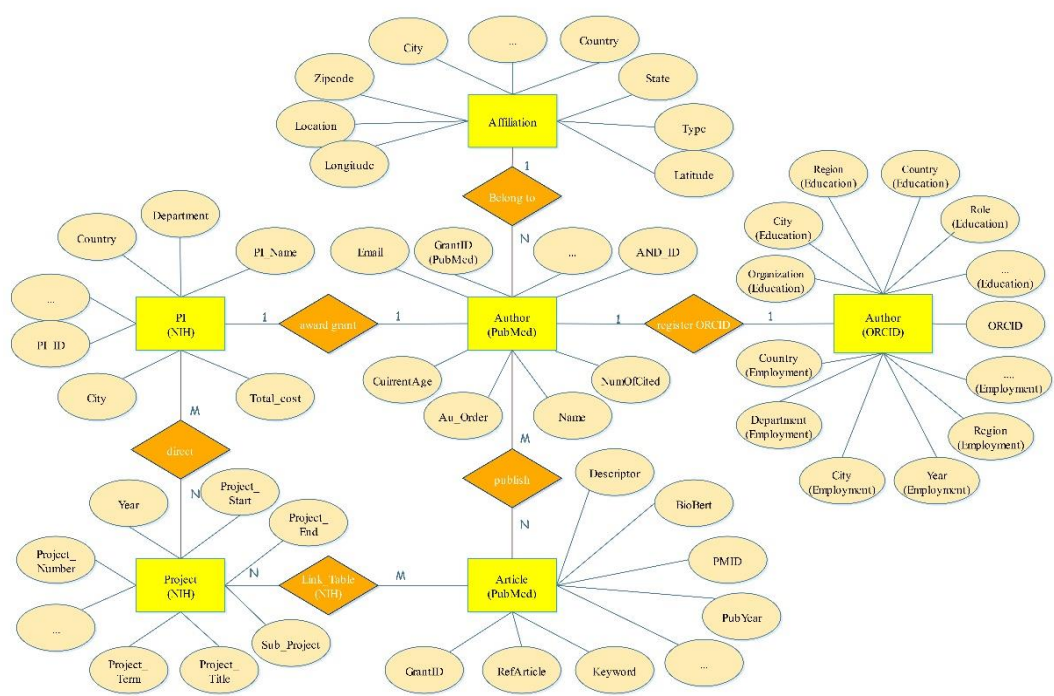

Figure 2. Entities and relationships in PKG

**(1) Project Data from NIH ExPORTER**

NIH ExPORTER provides data files that contains research projects funded by major funding agencies such as the Centers for Disease Control and Prevention (CDC), the NIH, the Agency for Healthcare Research and Quality (AHRQ), the Health Resources and Services Administration (HRSA), the Substance Abuse and Mental Health Services Administration (SAMHSA), and the U.S. Department of Veterans Affairs (VA). Furthermore, it provides publications and patents citing support from these projects. It consists of 49 data fields, including the amount of funding for each fiscal year, organization information of the PIs, and the details of the projects. According to our investigation, NIH-funded research accounts for 80.7% of all grants recorded in PubMed.

The NIH ExPORTER dataset contains a unique PI_ID for each scholar who received NIH funding between 1985 and 2018, and his or her PMIDs of the published articles. Through the mapping of PMIDs in NIH ExPORTER to PMIDs in PubMed, 1:N connections between the PI and the article have been established, paving the way for investigating the article details of a specific PI, and vice versa. Furthermore, by mapping PI names (last name, first initial, and affiliation) to author names that were listed in articles supported by the PI's projects, a 1:1 connection between the PI and the AND_ID was established, providing a way to obtain PI-related article information, regardless of whether the article was labeled with a project ID.

**(2) Employment History and Educational Background Data from ORCID**

According to its website, "ORCID is a nonprofit organization helping to create a world in which all who participate in research, scholarship, and innovation are uniquely identified and connected to their contributions and affiliations across disciplines, borders, and time"[36]. It maintains a registry platform for researchers to actively participate in identifying their own publications, information about formal employment relationships with organizations, and educational backgrounds. ORCID provides an open-access dataset called ORCID Public Dataset 2018[6], which contains a snapshot of all public data in the ORCID Registry associated with an ORCID record that was created or claimed by an individual as of October 1, 2018. The dataset includes 7,132,113 ORCID iDs, of which 1,963,375 have educational affiliations and 1,913,610 have employment affiliations.

As a result of the proliferation of ORCID identifiers, PubMed has used ORCID identifiers as alternative author identifiers since 2013[37]. Using the aforementioned two steps, we could map ORCID records to the PubMed authors. Our first step was to map the author instances in PubMed to an ORCID record based on the feature combinations of article DOI and author name (last name and first initial). Because the DOI is not a compulsory field for PubMed, we appended the feature combinations of article titles, journals, and author names to map the records between the two datasets. The result contained many 1:1 connections between a disambiguated author of PubMed and an ORCID record. Furthermore, 1:1 connections between AND_ID and ORCID iD, and 1:N connections between AND_ID and background information (education and employment) were established.

**(3) Fine-Grained Affiliation Data**

The MapAffil 2016 dataset[3] resolves PubMed authors' affiliation strings to cities and associated geocodes worldwide. This dataset was constructed based on a snapshot of PubMed (which included the Medline and PubMed-not-Medline records) acquired in the first week of October 2016. Affiliations were linked to a specific author on a specific article. Prior to 2014, PubMed only recorded the affiliation of the first author. However, MapAffil 2016 covered some PubMed records that lacked affiliations and were harvested elsewhere, such as from PMC, NIH grants, the Microsoft Academic Graph, and the Astrophysics Data System. All affiliation strings were processed using MapAffil to identify and disambiguate the most specific place names. The dataset provides the following fields: PMID, author order, last name, first name, year of publication, affiliation type, city, state, country, journal, latitude, longitude, and Federal Information Processing Standards (FIPs) code.

The MapAffil 2016 dataset does have a limitation because it does not cover the PubMed data after 2015 (covering 62.9% affiliation instances in PubMed). Consequently, we performed an additional step to improve the fraction of coverage. We collected authors (who published their first article before 2016 and continued publishing articles after 2015) by their AND_IDs. The new affiliation instances of the author after 2015 succeeded their corresponding fine-grained affiliation data from the affiliation instances before 2016 (fraction of affiliation instance coverage increased to 84.2%) if the author did not change affiliation. We also applied an up-to-date open-source library Affiliation Parser[4] to extract additional fine-grained affiliation fields from all affiliation instances, including department, institution, email, ZIP code, location, and country.

Table 2 summarizes the date coverage and version information of integrated datasets and open-access software used to extract data.

Table 2. Date coverage and version information of data sources

| Data Source | Start Year | End Year | Version Information |
|---|---|---|---|
| PubMed 2019 baseline files[39] | 1781 | 2018 | The PubMed 2019 baseline files were released in December 2018. It also includes 13,097 papers published after 2018 and majority of them are preprints. |
| Author-ity dataset[5] | 1865 | 2008 | The dataset was generated based on PubMed 2009 baseline |

| | | | files. It also includes AND results of 93,228 papers published after 2008, and majority of them are preprints. |
|---|---|---|---|
| Semantic Scholar dataset[9] | 1786 | 2019 | The dataset was released on January 31, 2019. |
| NIH ExPORTER dataset[10] | 1985 | 2018 | The articles marked with projects span from 1981 to 2018, and project details cover from 1985 to 2018. The dataset was downloaded in June 2018. |
| Employment History Data from ORCID[6] | 1913 | 2018 | The dataset was released on October 22, 2018. ORCID publishes the data once per year. |
| Educational Background Data from ORCID[6] | 1913 | 2018 | The dataset was released on October 22, 2018. ORCID publishes the data once per year. |
| MapAffil 2016 dataset[3] | 1975 | 2017 | The dataset is based on a snapshot of PubMed taken in the first week of October, 2016, and was released on April 5, 2018. |
| Affiliation Parser Library[4] | 1786 | 2019 | Fast and simple parser for MEDLINE and PubMed Open-Access affiliation string, which was published on March 15, 2018. We apply it to parse multiple fields from the affiliation string, including department, institution, zip code, location, and country. |

# Data Records

We built PKG with bio-entities extracted from PubMed abstracts, AND results of PubMed authors, and the integrated multi-source information. This dataset is freely available on Figshare[38]. It contains seven comma-separated value (CSV) files named "Author_List," "Bio_entities_Main," "Bio_entities_Mutation," "Affiliations," "Researcher_Employment," "Researcher_Education," and "NIH_Projects". The details are presented in Table 3. PubMed raw data are not included into Figshare file set because the amount of PubMed raw data is too large and they are not generated or altered by our methods. PubMed raw data can be freely downloaded from PubMed website[39]. We also provide download link (http://er.tacc.utexas.edu/datasets/ped), which contains both the PubMed raw data and PKG dataset to facilitate the application of PKG dataset.

Table 3. Dataset details.

| File | # of Lines | # of Distinct PMIDs | # of Distinct AND_IDs | Short description |
|---|---|---|---|---|
| Author_List | 114,345,178 | 28,510,300 | 14,830,461 | CSV file containing PubMed authors and AND_IDs. |
| Bio-entities_Main | 330,394,494 | 18,361,409 | - | CSV file containing all types of extracted bio-entities by BioBERT. |
| Bio-entities_Mutation | 1,388,341 | 312,099 | - | CSV file containing additional items of mutations from Bio-entities_Main file. |
| Affiliations | 46,065,099 | 19,601,383 | 8,300,984 | CSV file containing affiliations and their extracted fine-grained items. |
| Researcher_Employment | 532,356 | - | 276,483 | CSV file containing employment history from ORCID. |
| Researcher_Education | 512,267 | - | 268,610 | CSV file containing educational background from ORCID. |
| NIH_Porjects | 12,340,431 | 1,790,949 | 102,070 | CSV file containing projects from NIH ExPORTER and mapping relation between PI_ID, PMID, and AND_ID. |

Note: In file Author_List, about 1.3 million (1.15%) author instances cannot be disambiguated because they do not exist in Authority or Semantic Scholar dataset. Therefore, their AND_ID field values were set to zero.

The statistics of all five types of extracted entities are presented in Table 4.

Table 4. Statistics of extracted entities

| | Species | Disease | Gene / Protein | Drug/Chemical | Mutation |
|---|---|---|---|---|---|
| **Total number of extracted entities** | 65,737,425 | 98,865,897 | 81,035,640 | 83,367,191 | 1,388,341 |
| **Distinct PMIDs for each type** | 13,717,884 | 12,708,292 | 7,914,735 | 9,681,294 | 312,099 |
| **Distinct entities for each type** | 84,203 | 36,704 | 25,489 | 134,574 | 208,466 |

Each data field is self-explanatory by its name, and fields with the same name in other tables follow the same data format that can be linked across tables. Tables 5–11 illustrate the field name, format, and short description of fields for each data file listed in Table 3.

Table 5. Data type for records of Author_List.

| Index | Format | # of Lines with non-empty values | Short description |
|---|---|---|---|
| id | Integer | 114,345,178 | Unique ID for each author instance. |
| PMID | Integer | 114,345,178 | Unique ID assigned by PubMed to identify PubMed articles. |
| AND_ID | Integer | 109,245,192 | Unique author ID allocated by AND. |
| AuOrder | Integer | 114,345,178 | Author order of the current author in the author list of current articles. |
| LastName | String | 114,130,643 | Last name of the current author. |
| ForeName | String | 113,452,639 | First name of the current author. |
| Initials | String | 114,007,764 | Middle initials of the current author. |
| Suffix | String | 513,508 | Suffix name of the current author. |
| AuNum | Integer | 114,345,178 | Co-author number of the current articles. |
| PubYear | Integer | 114,345,178 | Publication year of the current article. |
| BeginYear | Integer | 109,245,192 | Begin year of the current author's first article. |

Table 6. Data type for records of Bio_Entities_Main.

| Index | Format | # of Lines with non-empty values | Short description |
|---|---|---|---|
| id | Integer | 330,394,594 | Unique ID for each bio-entity instance. |
| PMID | Integer | 330,394,594 | Unique ID assigned by PubMed to identify PubMed articles. |
| Start | Integer | 330,394,594 | Start position of mention in an abstract. |
| End | Integer | 330,394,594 | End position of mention in an abstract. |
| Mention | String | 330,394,594 | Entity mentioned in an abstract. |
| EntityID | Integer | 265,304,264 | Normalized entity ID. |
| Type | String | 330,394,594 | Enumerated type of entity; values include species, disease, gene, drug, and mutation. |

Table 7. Data type for records of Bio_Entities_Mutation.

| Index | Format | # of Lines with non-empty values | Short description |
|---|---|---|---|
| Main_id | Integer | 1,388,341 | Foreign key references from Bio-entities_Main (id). |
| Mention | String | 1,388,341 | Mutation entity mentioned in the abstract. |
| MutationType | String | 1,388,341 | Normalized entity ID. |
| NormalizedName | String | 1,388,341 | Enumerated type of entity; values include species, disease, gene, drug, and mutation. |

Table 8. Data type for records of Affiliations.

| Index | Format | # of Lines with non-empty values | Short description |
|---|---|---|---|
| id | Integer | 46,065,099 | Unique ID for each affiliation. |
| PMID | Integer | 46,065,099 | Unique ID assigned by PubMed to identify PubMed articles. |
| AuOrder | Integer | 46,065,099 | Author order of the current author in the author list of the current article. |
| AND_ID | Integer | 42,242,447 | Unique author ID allocated by AND. |
| AffiliationOrder | Integer | 46,065,099 | Affiliation order in the affiliation list of the current author. |
| Affiliation | String | 42,676,487 | Affiliation string. |
| Department | String | 29,438,469 | The department that the author belongs to. |
| Institution | String | 38,955,031 | The institution that the author belongs to. |
| Email | String | 8,092,262 | The author's email address. |
| ZipCode | String | 16,573,810 | The postcode of this affiliation. |
| Location | String | 42,590,482 | The address of the affiliation. |
| Country | String | 39,536,798 | The country that the author belongs to. |
| City | String | 32,151,044 | The city that the author belongs to. |
| State | String | 31,910,547 | The state that the author belongs to. |
| AffiliationType | String | 35,706,926 | Enumerated type of affiliation; values include COM, EDU, EDU-HOS, GOV, HOS, MIL, ORG, and UNK. |
| Latitude | Float | 36,371,281 | The latitude of the affiliation. |
| Longitude | Float | 21,679,300 | The longitude of the affiliation. |
| Fips | Integer | 8,727,595 | FIPS code of the county that includes the geocode. |

Table 9. Data type for records of Researcher_Employment.

| Index | Format | # of Lines with non-empty values | Short description |
|---|---|---|---|
| id | Integer | 532,356 | Unique ID for each scholar's employment instance. |
| AND_ID | Integer | 532,356 | Unique author ID allocated by AND. |
| ORCID | String | 532,356 | Unique researcher ID that distinguishes the researcher from others. |
| Department | String | 426,597 | The department which the researcher belongs to. |
| BeginYear | String | 487,183 | The beginning year of the researcher's employment. |
| Organization | String | 532,356 | The institution which the researcher belongs to. |
| City | String | 532,356 | The city where the researcher works. |
| Region | String | 363,066 | The region where the researcher works. |
| Country | String | 532,356 | The country where the researcher works. |
| Identifier | String | 392,562 | The identifier of an organization. |
| IdSource | String | 392,562 | The provider of an organizations' identifier. |
| EndYear | String | 251,826 | The end year of the researcher's employment. |

Table 10. Data type for records of Researcher_Education.

| Index | Format | # of Lines with non-empty values | Short description |
|---|---|---|---|
| id | Integer | 512,267 | Unique ID for each scholar's education instance. |
| AND_ID | Integer | 512,267 | Unique author ID allocated by AND. |
| ORCID | String | 512,267 | Unique researcher ID that distinguishes the researcher from others. |
| BeginYear | String | 453,122 | The beginning year of the researcher's education. |
| Organization | String | 512,267 | The organization the researcher has been educated. |
| City | String | 512,267 | The city where the researcher works. |
| Region | String | 378,188 | The region where the researcher works. |
| Country | String | 512,267 | The country where the researcher works. |
| Identifier | String | 410,239 | The identifier of an organization. |
| IdSource | String | 410,239 | The provider of an organizations' identifier. |
| EndYear | String | 440,750 | The end year of the researcher's education. |
| Role | String | 487,218 | The degree that the researcher received. |

Table 11. Data type for records of NIH_Projects.

| Index | Format | # of Lines with non-empty values | Short description |
|---|---|---|---|
| id | Integer | 12,340,431 | Unique ID for each project instance. |
| AND_ID | Integer | 11,013,198 | Unique author ID allocated by AND. |
| PI_ID | String | 12,340,431 | Unique PI ID allocated by NIH. |
| PMID | Integer | 12,340,431 | Unique ID assigned by PubMed to identify PubMed articles. |
| ProjectNumber | String | 12,340,431 | Project number of the current project. |
| subProjectNumber | String | 9,438,420 | Subproject number of the current project. |
| PI_Name | String | 12,340,431 | Full name of a PI. |

Updating PKG is a complex task because it is subject to the update of different data sources and requires significant computation. In the future, we hope to refresh PKG quarterly based on PubMed updated files and updated datasets from other sources. We may also develop an integrative ontology to integrate all types of entities.

# Technical Validation

**Validity of Bio-Entity Extraction**

To validate the performance of the bio-entity extraction, we established BERT and the state-of-the-art models as baselines. Then, we calculated the entity-level precision, recall, and F1 scores of these models as evaluation metrics. The datasets and the test results of biomedical NER are presented in Table 12.

Table 12. Test results of biomedical NER.

| Entity Type | Datasets | Metrics | State-of-the-art | BERT (Wiki + Books) | BioBERT (+ PubMed + PMC) |
|---|---|---|---|---|---|
| Disease | NCBI disease[40] | P % | 86.41 | 84.12 | **89.04** |
|  |  | R % | 88.31 | 87.19 | **89.69** |

|  |  | F % | 87.34 | 85.63 | **89.36** |
|  | 2010 i2b2/VA[41] | P % | <u>87.44</u> | 84.04 | **87.50** |
|  |  | R % | **86.25** | 84.08 | 85.44 |
|  |  | F % | **86.84** | 84.06 | <u>86.46</u> |
|  | BC5CDR[42] | P % | 85.61 | 81.97 | **85.86** |
|  |  | R % | 82.61 | 82.48 | **87.27** |
|  |  | F % | 84.08 | 82.41 | **86.56** |
| Drug/Chemical | BC5CDR[42] | P % | **94.26** | 90.94 | <u>93.27</u> |
|  |  | R % | 92.38 | 91.38 | **93.61** |
|  |  | F % | <u>93.31</u> | 91.16 | **93.44** |
|  | BC4CHEMD[43] | P % | 91.30 | 91.19 | **92.23** |
|  |  | R % | 87.53 | 88.92 | <u>90.61</u> |
|  |  | F % | 89.37 | 90.04 | **91.41** |
| Gene/Protein | BC2GM[44] | P % | 81.81 | 81.17 | **85.16** |
|  |  | R % | 81.57 | 82.42 | **83.65** |
|  |  | F % | 81.69 | 81.79 | **84.40** |
|  | JNLPBA[45] | P % | **74.43** | 69.57 | <u>72.68</u> |
|  |  | R % | **83.22** | 81.20 | <u>83.21</u> |
|  |  | F % | **78.58** | 74.94 | <u>77.59</u> |
| Species | LINNAEUS[46] | P % | <u>92.80</u> | 91.17 | **93.84** |
|  |  | R % | **94.29** | 84.30 | <u>86.11</u> |
|  |  | F % | **93.54** | 87.6 | <u>89.81</u> |
|  | Species 800[47] | P % | **74.34** | 69.35 | <u>72.84</u> |
|  |  | R % | <u>75.96</u> | 74.05 | **77.97** |
|  |  | F % | <u>74.98</u> | 71.63 | **75.31** |
| Average |  | P % | <u>85.38</u> | 82.61 | **85.82** |
|  |  | R % | <u>85.79</u> | 84.00 | **86.40** |
|  |  | F % | <u>85.53</u> | 83.25 | **86.04** |

In Table 12, we report the precision (P), recall (R), and F1 (F) scores of each dataset. The highest scores are in **boldface**, and the second-highest scores are <u>underlined</u>. Sachan *et al.* (2017) [48] reported the scores of the state-of-the-art models for the NCBI disease and BC2GM datasets, presented in Table 10. Moreover, the scores for the 2010 i2b2/VA dataset were obtained from Zhu *et al.* (2018)[49] (single model), and the scores for the BC5CDR and JNLPBA datasets were obtained from Yoon *et al.* (2018) [13]. The scores for the BC4CHEMD dataset were obtained from Wang *et al.* (2018) [50], and scores for the LINNAEUS and Species-800 datasets were obtained from Giorgi and Bader (2018)[51].

According to Table 12, BERT, which is pre-trained on the general domain corpus, was highly effective. On average, the state-of-the-art models outperformed BERT by 2.28% in terms of the F1 score. However, BioBERT obtained the highest F1 score in recognizing Genes/Proteins, Diseases, and Drugs/Chemicals. It outperformed the state-of-the-art models by 0.51% in terms of the F1 score, on average.

**Validity of Multi-type Entity Normalization**

We used the multi-type normalization model to assign unique IDs to synonymous entities. Table 13 presents the performance of the multi-type entity normalization model.

Table 13. Performance of the multi-type normalization model.

| Entity type | Normalization model | Test sets | Precision % | Recall % | F1 score % | Accuracy % |
|---|---|---|---|---|---|---|
| Gene/Protein | GNormPlus | BC2 Gene Normalization, human species[52] | 87.1 | 86.4 | 86.7 | - |
|  |  | BC3 Gene Normalization, multispecies[53] | - | - | 50.1 | - |
| Disease | Sieve-based entity linking | ShARe/CLEF eHealth Challenge corpus[54] | - | - | - | 90.75 |
|  |  | NCBI disease | - | - | - | 84.65 |
| Mutation | tmVar 2.0 | OSIRISv1.2[55] | 97.20 | 80.62 | 88.14 | - |
|  |  | Thomas[56] | 89.94 | 88.24 | 89.08 | - |
| Species | Dictionary lookup of | BioCreative III GN[58] | - | - | 46.91 | - |

| | SR4GN[57] | | | | | |

Note: There are empty cells in the table because GNormPlus and tmVar 2.0 did not report their accuracies, the sieve-based entity linking model only reported its accuracy, and SR4GN only reported its F1 score. The authors of tmChem did not report the normalization performance of tmChem independently, so there were no performance data for Drug/Chemical.

As shown in Table 13, with respect to genes and proteins, there were 75 different species in the BC3 Gene Normalization (BC3GN) test set, but GNormPlus focused only on seven of these species. Consequently, GNormPlus achieved a considerably lower F1 score of 36.6% on the multispecies test set (BC3GN) than on the human species test set (BC2GN). For mutations, tmVar 2.0 achieved F1 scores close to 90% on two corpora: OSIRISv1.2 and the Thomas corpus.

**Validity of Author Name Disambiguation**

The validation of author disambiguation remains a challenge because there is a lack of abundant validation sets. We applied a method using the NIH ExPORTER-provided information on NIH-funded researchers to evaluate the precision, recall, and F1 measures of the author disambiguation[59].

NIH ExPORTER provides information about the principal investigator ID (PI_ID) for each scholar who received NIH funding between 1985 and 2018. Because applicants established a unique PI_ID and used the PI_ID across all grant applications, these PI_IDs have extremely high fidelity. NIH ExPORTER also provides article PMIDs as project outputs, which can be conveniently used as a connection between PI_IDs and AND_ID.

We confirmed the bibliographic information of the NIH-funded scientists who received NIH funding during the years 1985–2018. Our AND evaluation steps were as follows: First, we collected project data for the years 1981–2018 in NIH ExPORTER, including 304,782 PI_ID records and the corresponding 331,483 projects. Secondly, we matched the projects to articles acknowledging support by the grant, which were also recorded in the NIH ExPORTER dataset. We matched 214,956 of the projects to at least one article and identified 1,790,949 articles funded by these projects. Some of these projects (116,527) did not match articles and were excluded. Because the NIH occasionally awards a project to a team that includes more than one PI, we eliminated the 13,154 records that contained multiple PIs because they could result in uncertain credit allocation. Consequently, our relevant set of PIs decreased to 147,027 individuals associated with 1,749,873 articles and 201,802 projects.

We then connected NIH PI_IDs from NIH ExPORTER to AND_IDs using the article PMIDs and author (PI)'s last name plus the initials as a crosswalk. This step resulted in 1,400,789 unique articles remaining, associated with 109,601 PI_IDs and 107,380 AND_IDs. Finally, we computed precision (P) based on the number of articles associated with the most frequent AND_ID-to-PI_ID matched over the number of all articles associated with a specific AND_ID[60]. Furthermore, we computed recall (R) based on the number of articles associated with the most frequent PI_ID-to-AND_ID matched over the number of all articles associated with a particular PI_ID[60]. Figure 3 summarizes the precision, recall, and F1 calculations.

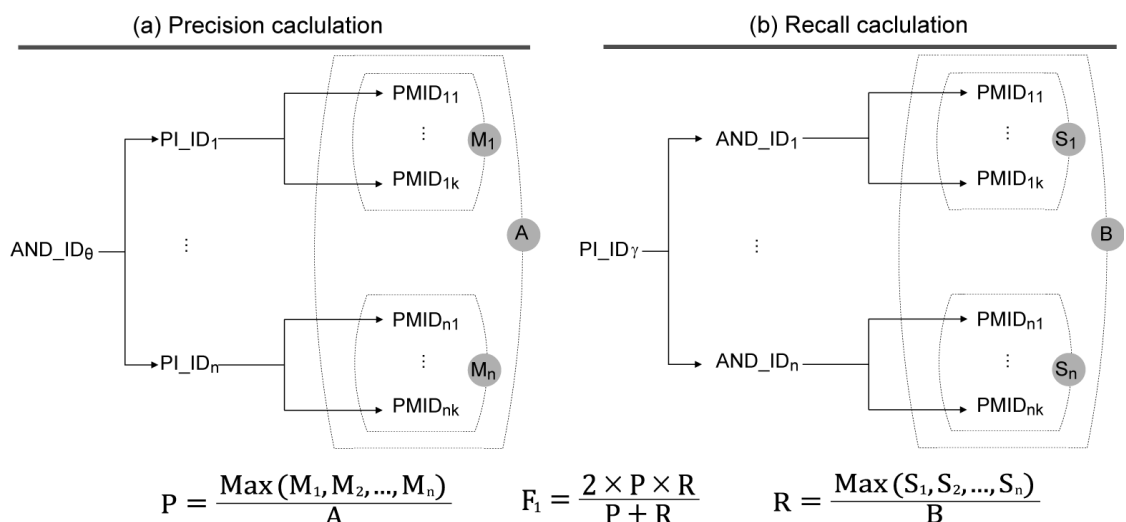

$$P = \frac{\text{Max}(M_1, M_2, ..., M_n)}{A} \qquad F_1 = \frac{2 \times P \times R}{P + R} \qquad R = \frac{\text{Max}(S_1, S_2, ..., S_n)}{B}$$

Figure 3. Calculation of Precision, Recall, and F1 Score

Table 14 illustrates the precision, recall, and F1 scores for Author-ity, Semantic Scholar, and our integrated AND result.

Table 14. Evaluation results of AND.

|  | Precision | Recall | F1 score |
|---|---|---|---|
| Author-ity | 99.43% | 96.92% | 98.16% |
| Semantic Scholar | 96.24% | 97.66% | 96.94% |
| AND Integration | 98.62% | 97.56% | 98.09% |

As presented in Table 14, after integrating the AND results of Author-ity and Semantic Scholar, we obtained a high-quality integrated AND result that outperformed Semantic Scholar by 1.15% in terms of the F1 score and had more comprehensive coverage (until 2018) than Author-ity (until 2009).

The evaluation results of AND might be slightly overestimated. The PIs of NIH grants usually have many publications over a long period and might be more likely to have rich information, such as affiliations and email addresses, about publications. Therefore, it should be easier to acquire higher performance on AND tasks than that of new entrants who published fewer papers and may lack of sufficient information for AND. Furthermore, approximately 1.15% of the author instances cannot be disambiguated since they do not exist in the Author-ity or Semantic Scholar AND results, which further slightly reduces the performance of AND results theoretically. However, the Semantic Scholar AND results and the AND Integration are evaluated based on the same baseline dataset with Author-ity in this section, and the evaluation of Author-ity performance using a random sample of articles indicates reliable high quality: the recall of the Author-ity dataset is 98.8%, the lumping (putting two different individuals into the same cluster) of the Author-ity dataset affects 0.5% of the clusters, and the splitting (assigning articles written by the same individual to more than one cluster) of the Author-ity dataset affects 2% of the articles[5]. Consequently, we believe these factors have a limited impact on AND performance.

## Usage Notes

Networking and collaboration have been associated with faculty promotions in academic medical centers[61]. Barriers exist for identifying researchers working on common bio-entities to facilitate collaboration. It is a challenge even at a single academic institution to identify potential collaborators who are working on the same bio-entities. This has led to many institution-specific projects profiling the faculty associated with the topics that they are studying.[62–65] The challenge is exacerbated when we search across multiple institutions.

Researchers, academic institutions, and the pharmaceutical industry often face the challenge of identifying researchers working on a specific bio-entity. A traditional bibliographic database specializes only in returning an enormous number of related articles for particular keyword or term searches. Bio-entity profiling for researchers offers an advantage over this traditional approach by identifying specific connections between bio-entities and disambiguated authors, in which bio-entity profiling for researchers can directly locate the core specialists whose research is focused on these bio-entities. Furthermore, a bipartite author-entity network projection analysis can identify a specific author's neighborhood with similar research interest, which is crucial for community detection and collaborative commendation.

We sought to use the PKG dataset to understand the trends over time of researcher-centric and bio-entity-centric activity by the following use cases: (1) researcher-centric for Stephen Silberstein, MD, a neurologist and expert in headache research; (2) calcitonin Gene-Related Peptide (CGRP), a target of inhibition for one of the newest therapeutics in migraine treatment; and (3) bipartite author-entity projection network analysis for coronavirus, a disease that causes respiratory illness with symptoms such as a fever, cough, and difficulty breathing.

For researcher-centric and bio-entity-centric activities, we collected 455 articles with Dr. Silberstein as an author and 7,877 articles on CGRP in the PKG dataset from 1970 to 2018 and extracted the bio-entities from these articles. Several publications and bio-entities were used for profiling the career of Dr. Silberstein. Several publications and the author's distribution were used for profiling CGRP. For bipartite author-entity projection network analysis, we collected 9,778 articles on coronavirus in the PKG dataset from 1969 to 2019.

**Researcher-Centric Activity**

For Dr. Silberstein, 539 bio-entities, including 342 diseases, 142 drugs, 24 genes, 17 species, and 14 mutations, were extracted from 455 articles. As depicted in Figure 4(a), "Headache" and "migraine" were his two most studied diseases, reaching 21 and 19 articles, respectively, in 2004. We trended his research over time on triptans, starting with sumatriptan. CGRP began to emerge in his publications starting in 2015. We noted the five researchers that have collaborated with Dr. Silberstein through his career and map with PKG their collaborations, interactions, and institutions over time. Visualizing the profiles of individual researchers can help to understand the trends in their topics of interest and collaboration patterns to enable an understanding of collaboration factors that may be associated with academic success or scientific discovery.

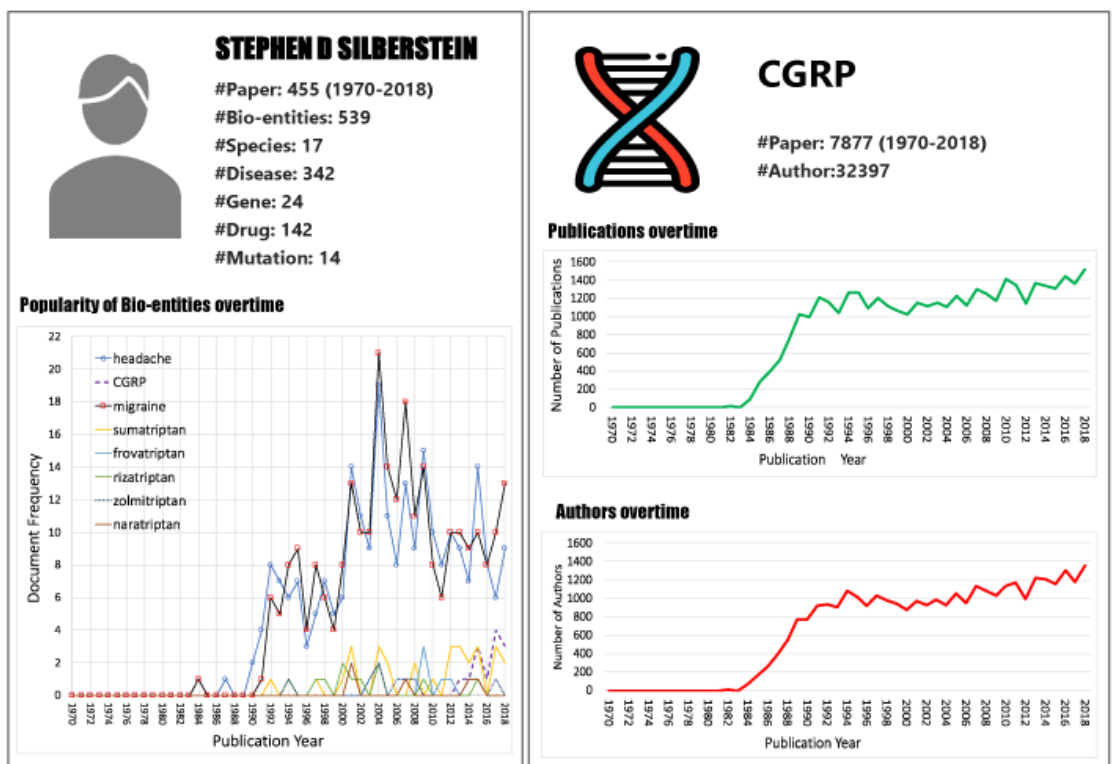

Figure 4. Trends over time of researcher-centric and bio-entity-centric activity

**Bio-Entity-Centric Activity**

For CGRP, there are currently 7,877 articles by 32,392 authors on CGRP dating back to 1982. Figure 4(b) illustrates that there was a dramatic increase in the number of CGRP-related articles, from 13 in 1982 to 1,209 in 1991, with a steady increase to 1,517 in 2018. The trend of the number of authors over time was similar to that of the volume of articles on CGRP.

As we demonstrated with a previous analysis of the repurposing of Aspirin[66-67], we observe research on CGRP starting at approximately the same time as the research on triptans for the treatment of migraines. Research on the pathophysiology of migraines identified a central role of the neuropeptide calcitonin gene-related peptide (CGRP), which is thought to be involved with the dilation of cerebral and dural blood vessels, release of inflammatory mediators, and the transmission of pain signals[68]. Research on the mechanism of the action of triptans—serotonin receptor agonists—has led to an understanding that they normalize elevated CGRP levels, which among other mechanisms, has led to an improvement in migraine headache symptoms. Consequently, papers in high-impact journals have called for identifying molecules and the development of drugs to directly inhibit CGRP[69], which has since led to the development of CGRP inhibitors as a new class of migraine treatment medications.

**Bipartite Author-Entity Network**

A total of 28,223 disambiguated authors and 5,379 distinct bio-entities of coronavirus articles were used to construct author-bio-entity bipartite network. Figure 5 illustrated the bipartite network (Figure 5(a)) and its author projection (Figure 5(b)) and bio-entity projection (Figure 5(c)). In Figure 5(a), the author vertices are blue, and the bio-entity vertices are pink. A link between a bio-entity and an author exists if and only if this bio-entity has been researched by that author. Connections between two authors or between two bio-entities are not allowed. The edge weight is set as the number of papers an author published that mention a bio-entity. In Figure 5(b) and 5(c), the edge weight is set as the number of common neighbors for the author and bio-entity, respectively. Vertices are marked with different colors to show their community attribution.

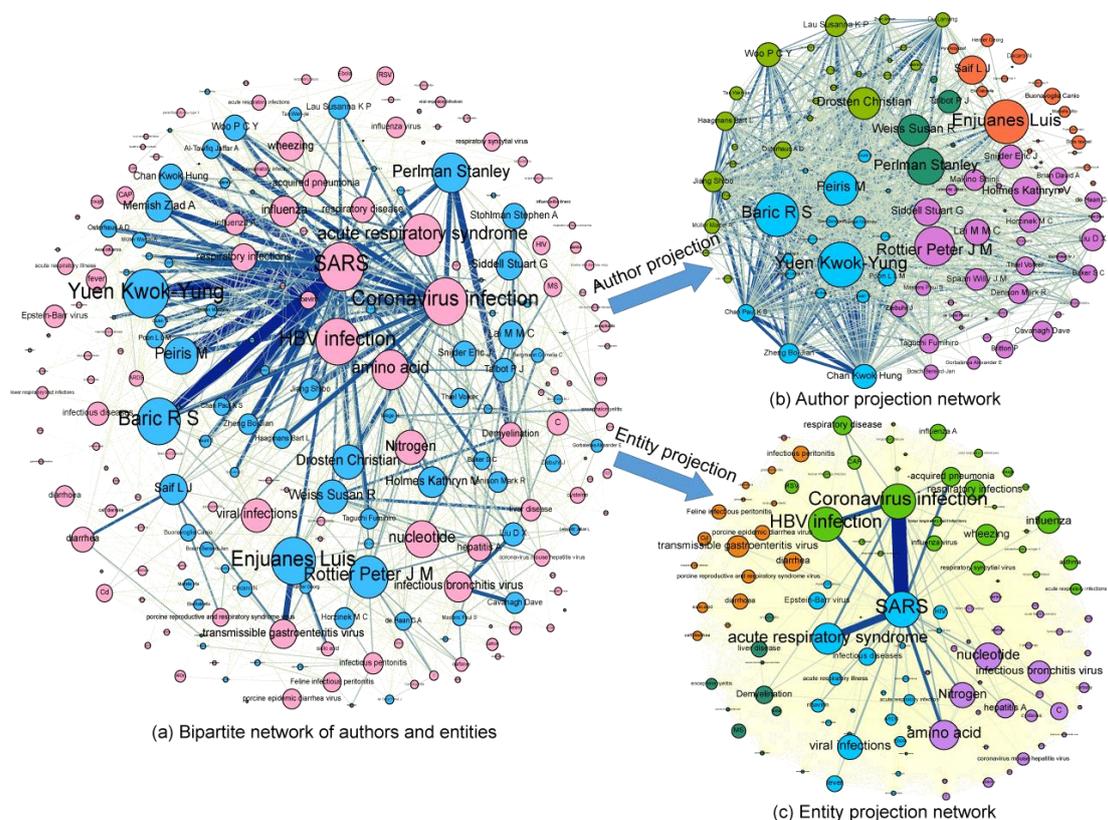

Figure 5. Bipartite network analysis of coronavirus

Figure 5(a) illustrates a distinct relationship between authors and their focused bio-entities. For example, the disease SARS have been frequently studied by author Baric R S, Yuen Kwok-Yung, and Zheng Bo-Jian. In addition to SARS, Baric R S is also interested in coronavirus infection and HBV infection. Figure 5(b) depicts the common research interest relationship between authors. Strong connections between authors may indicate that they collaborated multiple times, such as Chan Kwok Hung and Yuen Kwok-Yung, who published 69 papers together. These connections may also indicate author pairs that have similar research interests but never collaborated, such as Baric R S and Yuen Kwok-Yung, which is crucial for the collaborative commendation. Similarly, those connections between bio-entities in Figure 5(c) indicate that they have been studied by authors with similar research interests, which can be further applied to discover the hidden relations between bio-entities.

## Code Availability

We have made the pre-trained weights of BioBERT freely available at https://github.com/naver/biobert-pretrained, and the source code for fine-tuning BioBERT available at https://github.com/dmis-lab/biobert.

## Acknowledgments

This work was supported by National Social Science Fund of China [18BTQ076], Chinese National Youth Foundation Research [61702564], Natural Science Foundation of Guangdong Province [2018A030313981], Soft Science Foundation of Guangdong Province [2019A101002020], National Research Foundation of Korea [NRF-2019R1A2C2002577] and [NRF-2017R1A2A1A17069645], and US National Institutes of Health [P01AG039347]. The authors acknowledge the Texas Advanced Computing Center (TACC) at The University of Texas at Austin for providing storage resources that have contributed to the research results reported within this paper. URL: http://www.tacc.utexas.edu.

## Author contributions

YD, JX, and DL proposed the idea and supervised the project.
JX, YD and MS wrote and revised this manuscript.
SK, MJ, DK, and JK conducted the bio-entity extraction and validity.
JR, XL, WX, YB, CC, and IAE conducted the usage notes.
VIT and MS conducted the author name disambiguation and validity.

## Competing interests

The authors declare that they have no competing interests with respect to this paper.